\documentclass[pra,twocolumn,superscriptaddress]{revtex4-2}
\usepackage{graphicx,empheq,amssymb,scalerel,tikz}
\usepackage[colorlinks=true, citecolor=blue, urlcolor=blue, linkcolor=blue ]{hyperref}
\usetikzlibrary{svg.path}
\definecolor{orcidlogocol}{HTML}{A6CE39}
\tikzset{
	orcidlogo/.pic={
		\fill[orcidlogocol] svg{M256,128c0,70.7-57.3,128-128,128C57.3,256,0,198.7,0,128C0,57.3,57.3,0,128,0C198.7,0,256,57.3,256,128z};
		\fill[white] svg{M86.3,186.2H70.9V79.1h15.4v48.4V186.2z}
		svg{M108.9,79.1h41.6c39.6,0,57,28.3,57,53.6c0,27.5-21.5,53.6-56.8,53.6h-41.8V79.1z M124.3,172.4h24.5c34.9,0,42.9-26.5,42.9-39.7c0-21.5-13.7-39.7-43.7-39.7h-23.7V172.4z}
		svg{M88.7,56.8c0,5.5-4.5,10.1-10.1,10.1c-5.6,0-10.1-4.6-10.1-10.1c0-5.6,4.5-10.1,10.1-10.1C84.2,46.7,88.7,51.3,88.7,56.8z};}}
\newcommand\orcid[1]{\href{https://orcid.org/#1}{\mbox{\scalerel*{\begin{tikzpicture}[yscale=-1,transform shape]\pic{orcidlogo};\end{tikzpicture}}{|}}}}

\begin{document}
\title{Engineering second-order nodal-line semimetals by breaking $\mathcal{PT}$ symmetry and periodic driving}
\author{Ming-Jian Gao}
\affiliation{Lanzhou Center for Theoretical Physics, Key Laboratory of Theoretical Physics of Gansu Province, Lanzhou University, Lanzhou 730000, China}
\author{Hong Wu\orcid{0000-0003-3276-7823}}
\affiliation{Lanzhou Center for Theoretical Physics, Key Laboratory of Theoretical Physics of Gansu Province, Lanzhou University, Lanzhou 730000, China}
\author{Jun-Hong An\orcid{0000-0002-3475-0729}}
\email{anjhong@lzu.edu.cn}
\affiliation{Lanzhou Center for Theoretical Physics, Key Laboratory of Theoretical Physics of Gansu Province, Lanzhou University, Lanzhou 730000, China}
\begin{abstract}
Hosting unique drumhead surface states enclosed by nodal lines, topological nodal-line semimetals exhibit novel transport phenomena. Thus, the exploration of topological semimetals with different nodal-line structures has attracted much attention. In this paper, we first find a second-order nodal line semimetal (SONLS), which has coexisting hinge Fermi arcs and drumhead surface states, in a $\mathcal{PT}$-symmetry broken system. Then, without changing the intrinsic parameters, we artificially create exotic hybrid-order nodal-line semimetals hosted by different quasienergy gaps and rich nodal-line structures including nodal chains, crossing ring nodal nets, crossing line nodes, and nodal nets by applying a periodic driving on our SONLS. Enriching the classification of topological semimetals, such Floquet engineered high tunability of the orders and nodal-line structures of the SONLS sets up a foundation for exploring its further applications.
\end{abstract}
\maketitle

\section{Introduction}
The rapid development of topological phases \cite{RevModPhys.82.3045, RevModPhys.83.1057, RevModPhys.88.035005, RevModPhys.90.015001, RevModPhys.93.025002} not only enriches the paradigm of condensed matter physics, but also has a profound impact on many classical systems, such as phononic and photonic crystals \cite{RevModPhys.91.015006,RN4,Wei2021}, electric-circuit arrays \cite{10.1093/nsr/nwaa065}, and mechanical systems \cite{doi:10.1126/science.aab0239}. In the topological phase family, topological semimetals \cite{RevModPhys.93.025002} have attracted much attention recently. From a fundamental physics perspective, their discovery successfully generalizes topological phases from bulk gapped systems to bulk gapless ones. From an application perspective, topological semimetals have exhibited an ability in designing topological devices by using their chiral-anomaly-induced giant magnetoresistance and high carrier mobility \cite{Jia2016,doi.org/10.1002/adma.201606202,Wang2017,Han2020}.

Symmetries play a leading role in classifying topological phases. Various novel topological semimetals, i.e., Dirac \cite{doi:10.1126/science.1245085,RN11,PhysRevLett.113.027603,doi:10.1126/science.1256742,PhysRevLett.108.140405,PhysRevLett.115.176404,RN12,PhysRevLett.115.126803,PhysRevLett.119.026404,RN5}, Weyl \cite{PhysRevB.83.205101,doi:10.1126/science.aaa9297,PhysRevX.5.031013,PhysRevX.5.031023,PhysRevLett.124.236401,RN2,RN3,RN9,RN10,PhysRevLett.125.146401,PhysRevLett.125.266804}, and nodal-line \cite{PhysRevB.96.041103,PhysRevLett.120.146602,PhysRevB.97.161111,PhysRevB.93.121113,PhysRevLett.115.036807,PhysRevB.93.205132,PhysRevB.99.041301} semimetals, have been found by exploring different symmetries. Dirac semimetals are present in systems with spatial inversion $\mathcal{P}$ and time-reversal $\mathcal{T}$ symmetries. If either of the two symmetries is broken, then Weyl semimetals may be formed. Nodal-line semimetals are present in systems with either $\mathcal{P}$ and $\mathcal{T}$ symmetries or mirror symmetry \cite{PhysRevB.104.235136}. Recently, a novel semimetal called a second-order nodal-line semimetal (SONLS), which has coexisting hinge Fermi arcs and drumhead surface states, was proposed in a $\mathcal{PT}$-invariant system \cite{PhysRevLett.125.126403,PhysRevLett.127.076401,PhysRevLett.128.026405}. From the viewpoint of the classification of topological phases, an important question is whether $\mathcal{PT}$ symmetry is an essential prerequisite for SONLS.

On the other hand, nodal-line semimetals hosting drumhead surface states enclosed by different nodal-line structures exhibit novel transport features, which builds a foundation of their measurement and application \cite{RN33,PhysRevLett.121.166802,PhysRevLett.122.196603,PhysRevLett.124.056402}. Various types of nodal-line structures, e.g., nodal rings \cite{PhysRevB.92.045108,RN32,PhysRevB.92.045126,PhysRevLett.118.016401}, nodal links \cite{PhysRevB.96.041103}, crossing line nodes \cite{PhysRevB.95.245208}, nodal chains \cite{RN33}, nodal knots \cite{PhysRevB.96.201305}, and nodal nets \cite{PhysRevLett.120.026402}, have been discovered in different systems. In practical applications, it is always desirable for transport features caused by different nodal-line structures to be tuned on demand. This is difficult to realize in static systems since their properties can no longer be changed once their material samples are fabricated. Coherent control via the periodic driving of external fields, called Floquet engineering, has become a versatile tool in creating novel topological phases \cite{PhysRevB.93.184306,PhysRevA.100.023622,PhysRevB.102.041119,PhysRevB.103.L041115,PhysRevLett.121.036401,PhysRevLett.123.016806,PhysRevLett.124.057001,PhysRevLett.124.216601,PhysRevB.103.L041115,PhysRevB.103.115308,PhysRevB.104.205117}. A natural question is whether we can realize a free tunability and conversion of the nodal-line structures and the topological phases of a SONLS by Floquet engineering.

\begin{figure}[tbp]
\centering
\includegraphics[width=\columnwidth]{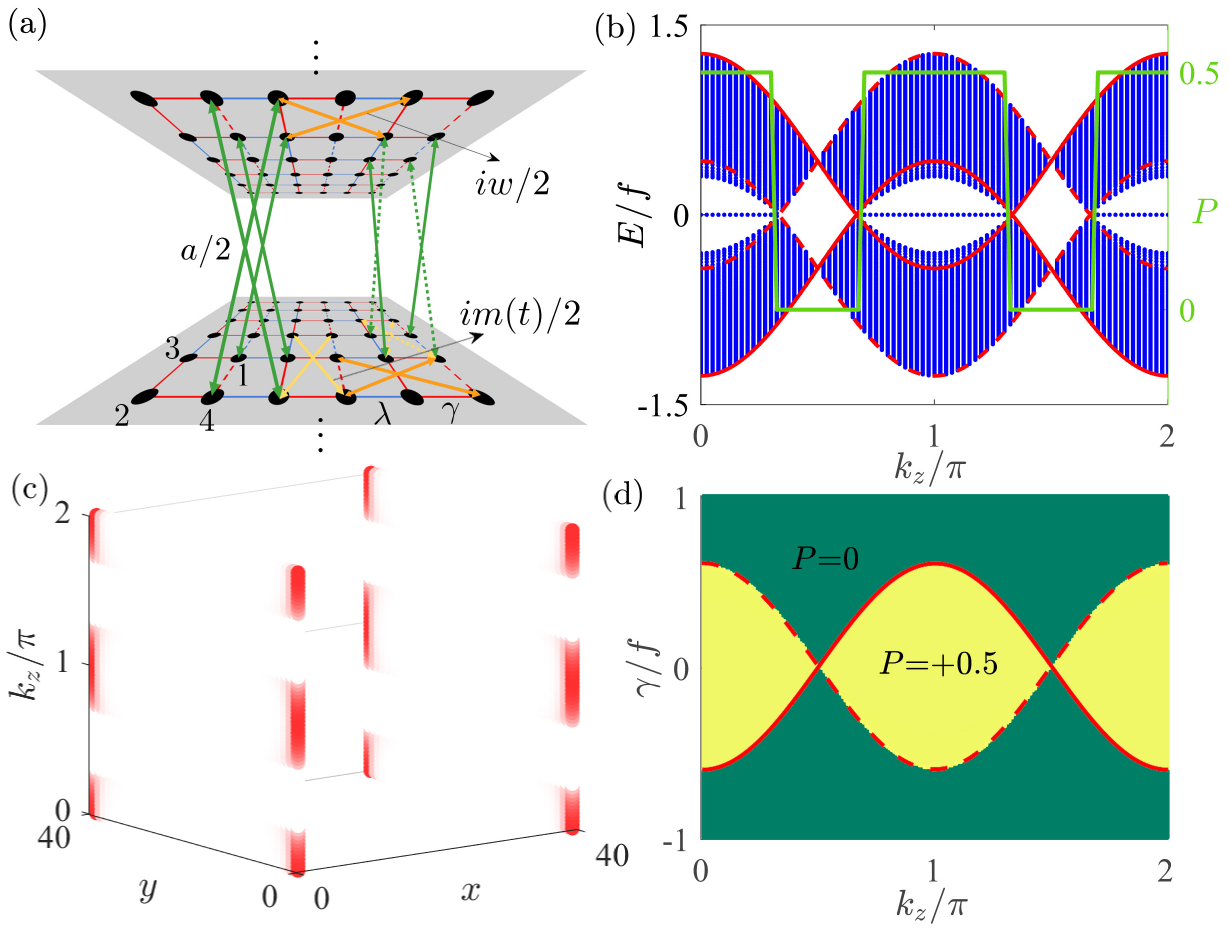}
\caption{(a) Schematic diagram of SONLS, with $\gamma$ and $\lambda$ being the intracell and nearest-neighbor intercell hopping rates, $im/2$ and $iw/2$ being the next-nearest-neighbor intercell ones, and $a/2$ being the interlayer one. The dashed lines denote the hopping rates with a $\pi$-phase difference from their solid lines. (b) Energy spectrum of $\mathcal{H}_{0}(\textbf{k})$ (blue lines) and quadrupole moment $P$ (green line) and (c) hinge Fermi arc as a function of $k_{z}$. The red solid (dashed) line is the dispersion relation along the high-symmetry line $k = 0$ ($\pi$). (d) Phase diagram described by $P$. We use $\gamma$ = $0.3f$, $a$ = $0.6f$, $\lambda$ = 0, and the lattice number $N_{x}$ = $N_{y}$ = 40.}\label{fig:1}
\end{figure}

Addressing these questions, we investigate the SONLS and its Floquet engineering. We discover a way to realize the SONLS in $\mathcal{PT}$-symmetry broken systems. It is found that a Dirac semimetal transforms into a SONLS when a perturbation breaking the $\mathcal{PT}$ symmetry is applied. We also explore the tunability of the nodal-line structures and the topological phases of the SONLS by Floquet engineering. Hybrid-order nodal-line semimetal and rich nodal-line structures including nodal chains, crossing ring nodal net, crossing line nodes, and nodal nets are created easily by applying periodic driving. The interconversion of different orders of topological semimetals is also realized by periodic driving. Enriching the family of topological phases, our result is helpful to explore their application.

\section{SONLS in $\mathcal{PT}$-symmetry broken system}
We consider a system of spinless fermions moving on a three-dimensional (3D) lattice [see Fig. \ref{fig:1}(a)]. Its Hamiltonian reads $\hat{H}_0=\sum_{\textbf{k}}$$\hat{C}^{\dagger}_{\textbf{k}}$$\mathcal{H}$(\textbf{k})$\hat{C}_{\textbf{k}}$ with $\hat{C}^{\dagger}_{\textbf{k}}$ = ($\hat{C}^{\dagger}_{\textbf{k},1}$ $\hat{C}^{\dagger}_{\textbf{k},2}$ $\hat{C}^{\dagger}_{\textbf{k},3}$ $\hat{C}^{\dagger}_{\textbf{k},4}$) and
\begin{equation} \label{text}
\begin{split}
\mathcal{H}_0(\textbf{k})&=[\gamma+\chi(k_{z})\cos k_{x}]\Gamma_{5}-\chi(k_{z})\sin k_{x}\Gamma_{3}\\
&+\,[\gamma+\chi(k_{z})\cos k_{y}]\Gamma_{2}+\chi(k_{z})\sin k_{y}\Gamma_{1},
\end{split}
\end{equation}
where $\chi(k_{z})=\lambda+a\cos k_{z}$, $\gamma$ is the intracell hopping rate, $\lambda$ is the intercell hopping rate, $a$ is the interlayer hopping rate between the nearest-neighbor unit cells, $\Gamma_{i}=\tau_{y}\sigma_{i}$ ($i=1,2,3$), and $\Gamma_{5}=\tau_{x}\sigma_{0}$, with $\tau_{i}$ and $\sigma_{i}$ being Pauli matrices, and $\tau_{0}$ and $\sigma_{0}$ being identity matrices. The system is a 3D generalization of the Benalcazar-Bernevig-Hughes model \cite{PhysRevB.96.245115} by considering interlayer hopping \cite{PhysRevLett.125.266804}. It is a Dirac-type second-order topological semimetal, which is sliced into a family of two-dimensional (2D) $k_z$-dependent second-order topological insulators and normal insulators separated by discrete Dirac points. The system has chiral symmetry $\mathcal{S} = \tau_{z}\sigma_{0}$, time-reversal symmetry $\mathcal{T} = K$, with $K$ being the complex conjugation, and the spatial inversion symmetry $\mathcal{P}$ = $\tau_{0}\sigma_{y}$. Thus its $\mathcal{PT}$ symmetry satisfies $\mathcal{(PT)}^{2}=-1$, which is different from the case of $\mathcal{(PT)}^{2}=1$ in Refs. \cite{PhysRevLett.125.126403,PhysRevLett.127.076401}. It also possesses mirror-rotation symmetries $\mathcal{M}_{x}$ = $\tau_{x}\sigma_{z}$, $\mathcal{M}_{y}$ = $\tau_{x}\sigma_{x}$, and $\mathcal{M}_{xy}=[(\tau_0+\tau_z)\sigma_x-(\tau_z-\tau_0)\sigma_z]/2$. So the topological phases are described by the Hamiltonian along the high-symmetry line $k_x=k_y\equiv k$, which is diagonalized into $\text{diag}[\mathcal{H}_0^{+}(k,k_z),\mathcal{H}_0^{-}(k,k_z)]$ with $\mathcal{H}^{\pm}_0(k,k_z)=\mathbf{h}^\pm\cdot{\pmb \sigma}$ and $\mathbf{h}^\pm=\sqrt{2}[\gamma+\chi(k_z)\cos k,\pm\chi(k_z)\sin k,0]$. A series of Dirac points satisfying $|\gamma|=|\chi(k_z)|$ when $k=0$ or $\pi$ are discovered, at which a $k_z$-dependent topological phase transition occurs. This is confirmed by the energy spectrum in Fig. \ref{fig:1}(b), where fourfold degenerate zero-mode states are formed. Their probability distributions in Fig. \ref{fig:1}(c) reveal that they are corner states in the sliced 2D space and form the hinge Fermi arcs of a 3D second-order topological semimetal. These corner states are topologically described by the $k_{z}$-dependent quadrupole moment \cite{PhysRevB.103.L041115,PhysRevB.100.245134,PhysRevB.100.245135}
\begin{equation}
P=\Big[\frac{\text{Im}\ln\det\mathcal{U}}{2\pi}-\sum_{{\bf n},i;{\bf m},j} \frac{X_{{\bf n},i;{\bf m},j}}{2L_xL_y}\Big]\text{mod}~1.\label{polrz}
\end{equation}
Here, the elements of $\mathcal{U}$ read $\mathcal{U}_{ab}\equiv \langle \psi_a|e^{i2\pi X/(L_xL_y)}|\psi_b\rangle$, $|\psi_\alpha\rangle$ ($\alpha=a,b$) satisfying $\hat{H}_0|\psi_a \rangle=E_{a}|\psi_a\rangle$ and $E_\alpha<0$ are the occupied eigenstates, and the coordinate $X_{{\bf n},i;{\bf m},j}=n_xn_y\delta_{{\bf n}{\bf m}}\delta_{ij}$ with $i,j=1,\cdots,4$ being the sublattices and $n_{x,y}$ being the numbers of unit cells. $P = 0.5$ signifies the corner states [see Figs. \ref{fig:1}(b) and \ref{fig:1}(d)].

\begin{figure}[tbp]
\centering
\includegraphics[width=\columnwidth]{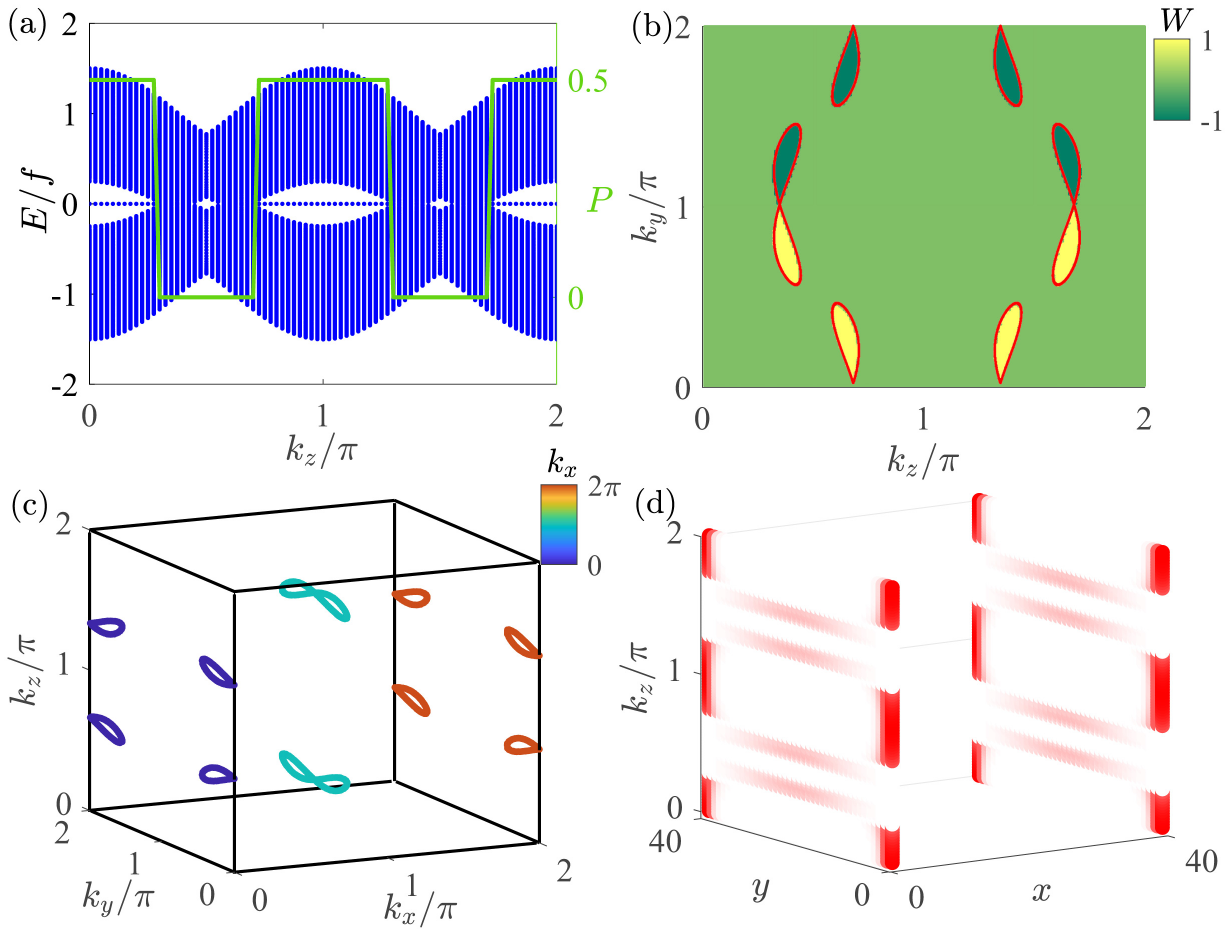}
\caption{(a) Energy spectrum and quadrupole moment (green line) of $\mathcal{H}_0$ + $\Delta\mathcal{H}$. (b) Winding number $W(k_y,k_z)$ and nodal lines (in red lines) in the $k_y$-$k_z$ plane. (c) Distribution of the nodal lines in the Brillouin zone. (d) Coexistence of the hinge Fermi arcs and the drumhead surface states. We use $m$ = $0.35f$, $w$ = $0$, and other parameters being the same as in Fig. \ref{fig:1}.}\label{fig:2}
\end{figure}

To generate a SONLS, we add a perturbation to Eq. \eqref{text}. Different from the utilization of a $\mathcal{PT}$-invariant perturbation \cite{PhysRevLett.125.126403,PhysRevLett.127.076401}, we generate a SONLS by a $\mathcal{PT}$-symmetry broken perturbation
\begin{equation}\label{per}
\Delta\mathcal{H}({\bf k}) = w\sin k_{x}\tau_{x}\sigma_{x} + m\sin k_{y}\tau_{x}\sigma_{z},
\end{equation}
where $w$ and $m$ are the $x$- and $y$-direction next-nearest-neighbor intercell hopping rates [see Fig. \ref{fig:1}(a)]. Equation \eqref{per} breaks time-reversal symmetry $\mathcal{T}$ and mirror-rotation symmetry $\mathcal{M}_{xy}$, while preserving chiral symmetry $\mathcal{S}$, spatial inversion symmetry $\mathcal{P}$, and mirror-rotation symmetries $\mathcal{M}_{x}$ and $\mathcal{M}_{y}$ of $\mathcal{H}_0(\textbf{k})$. Equation \eqref{per} exhausts all the possibilities in our system satisfying the above symmetry requirement. Other forms proportional to $\tau_i\sigma_j$ either cannot satisfy the symmetry requirements or cannot be realized in the lattice in Fig. \ref{fig:1}(a). We plot the energy spectrum under an open boundary condition in Fig. \ref{fig:2}(a). It is found that each Dirac point in Fig. \ref{fig:1}(b) spreads into a nodal line. The quadrupole moment $P=0.5$ guarantees that the corner nature of the zero-mode state is preserved. This proves the formation of a 2D sliced second-order topological phase. It is interesting to find that the regimes within the nodal-line loops are first-order surface flat bands. Such first-order topology is characterized by the winding number \cite{PhysRevLett.123.246801,RevModPhys.88.035005}
\begin{equation}
\begin{aligned}
W(k_y,k_z)=\frac{1}{4\pi i}\int_{-\pi}^{\pi} \textrm{Tr}[\mathcal{S}\mathcal{Q}(\textbf{k})\partial_{k_{x}} \mathcal{Q}(\textbf{k})]dk_x,
\end{aligned}
\end{equation} where $
\mathcal{Q}(\textbf{k})=\sum_{l=1,2}[|u_{-l}(\textbf{k})\rangle \langle u_{-l}(\textbf{k})|-|u_{l}(\textbf{k})\rangle \langle u_{l}(\textbf{k})|]$, with $|u_{l}(\textbf{k})\rangle$ satisfying $[\mathcal{H}_{0}(\textbf{k})+\Delta\mathcal{H}(\textbf{k})]|u_{l}(\textbf{k})\rangle = E_{l}(\textbf{k})|u_{l}(\textbf{k})\rangle$. The chiral symmetry makes $E_{-l}({\bf k})=-E_l({\bf k})$. We see from Fig. \ref{fig:2}(b) that the regimes with $W=\pm1$ exactly match with the regimes in the presence of the nodal lines. This verifies the first-order nature of the surface flat band within the formed nodal-line loops. The distribution of the nodal lines in the Brillouin zone in Fig. \ref{fig:2}(c) indicates that they reside in the planes of $k_x=0$ and $\pi$. The probability distribution of all the zero-mode states in Fig. \ref{fig:2}(d) exhibits hinge Fermi arcs in the 2D second-order topological insulator regimes and the drumhead surface states in the first-order surface flat-band regime are separated by nodal lines. This confirms the formation of the SONLS. Thus, different from Refs. \cite{PhysRevLett.125.126403,PhysRevLett.127.076401}, we find a way to realize SONLS in $\mathcal{PT}$-symmetry broken systems.

\begin{figure}[tbp]
\centering
\includegraphics[width=\columnwidth]{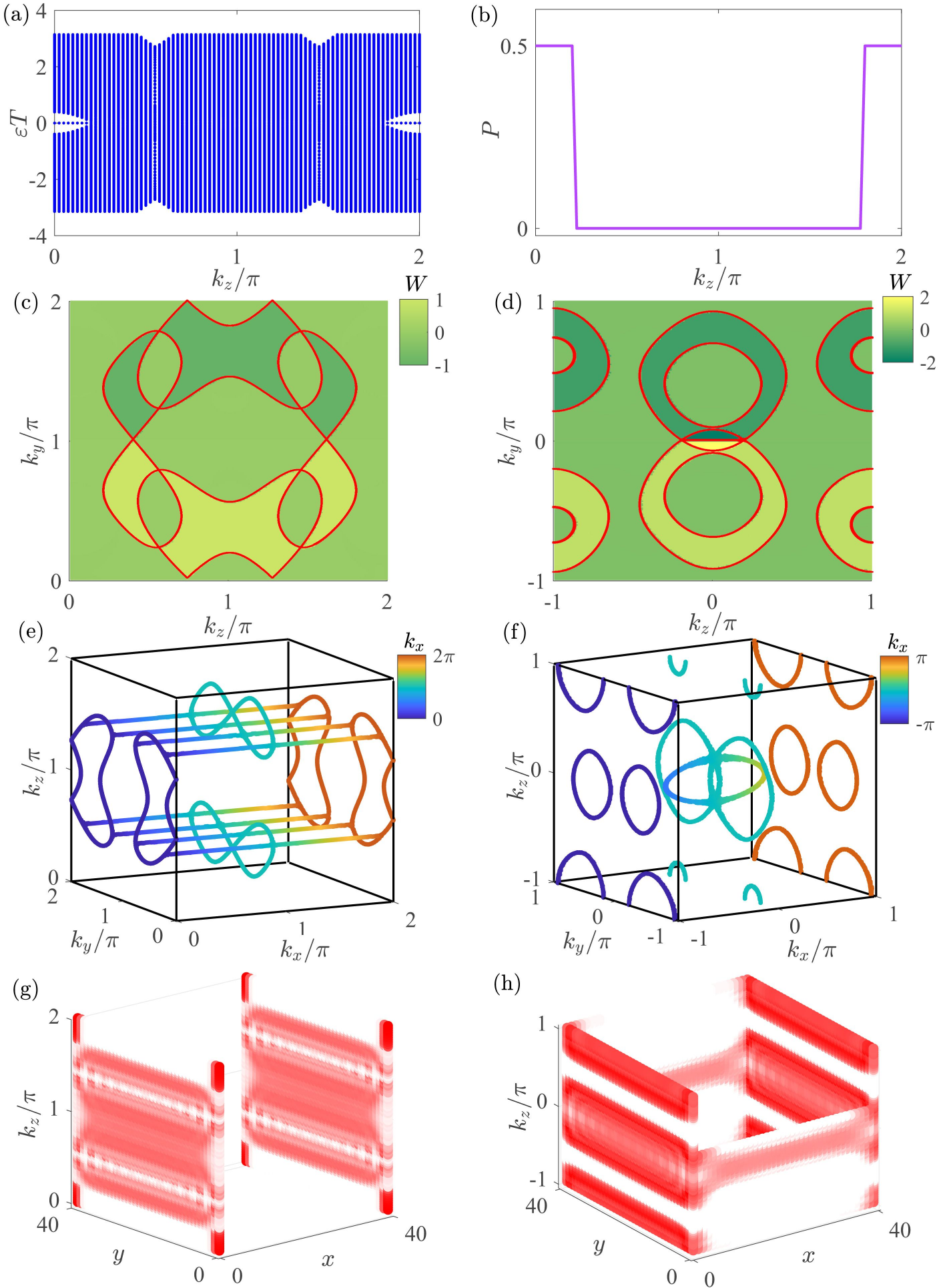}
\caption{(a) Quasinergy spectrum and (b) quadrupole moment of the periodically driven system. Winding numbers (c) $W_0$ and (d) $W_{\pi/T}$ (d), with the red lines denoting the projections of the nodal line on the $k_{y}$-$k_{z}$ plane. Nodal-line distribution of the (e) zero-mode and (f) $\pi/T$-mode in the Brillouin zone. (g) Hybrid-order nodal-line semimetal with coexisting second-order hinge Fermi arcs and first-order drumhead surface states in the zero mode and (h) purely first-order drumhead surface states in the $\pi/T$ mode. We use $\gamma=0.3f$, $a=0.6f$, $\lambda=0.1f$, $m_{1}=0$, $m_{2}=1.1$, $T_{1}=1.0f^{-1}$, and $T_{2}=1.5f^{-1}$.}\label{fig:3}
\end{figure}
\section{Floquet engineering to SONLS}
As they are determined by the hopping rates, the topological features and the nodal-line structure of any SONLS in a static system cannot be changed once their material is fabricated. Without resorting to changing the intrinsic parameters, we use Floquet engineering to realize such controllability. We consider that $m$ in Eq. \eqref{per} is periodically driven as
\begin{equation}\label{m0}
\begin{split}
m(t)= \left \{
 \begin{array}{ll}
m_{1}f,                    & t\in [nT,nT+T_{1})\\
m_{2}f,                    & t\in [nT+T_{1},(n+1)T)
 \end{array}
 \right.,
 \end{split}
 \end{equation}
where $n \in$ $\mathbb{Z}$, $T$ = $T_{1}$ + $T_{2}$ is the driving period, and $f$ is an energy scale to make the driving amplitudes $m_{j}$ ($j=1,2$) dimensionless. Such driving has been used in generating a time crystal in a spin chain system \cite{Choi2017,Zhang2017} and simulating large-Chern-number Floquet Chern insulators \cite{yang2022observation,PhysRevB.93.184306} in the nitrogen-vacancy center system. The periodic system does not have an energy spectrum because its energy is not conserved. However, the Floquet theorem reveals that $|u_\alpha(t)\rangle$ and $\varepsilon_\alpha$ in the Floquet equation $[\hat{H}(t)-i\partial_t]|u_\alpha(t)\rangle=\varepsilon_\alpha|u_\alpha(t)\rangle$ play the same roles as the stationary states and eigen energies in static systems. They are thus called quasistationary states and quasienergies, respectively \cite{PhysRevA.7.2203,PhysRevA.91.052122}. The topology of our periodic system is defined in the quasienergy spectrum. It can be proven that the Floquet equation is equivalent to $\hat{U}_T|u_\alpha(0)\rangle=e^{-i\varepsilon_\alpha T}|u_\alpha(0)\rangle$, where $\hat{U}_T=\mathbb{T}e^{-i\int_0^T\hat{H}(t)dt}$ is a one-period evolution operator with $\mathbb{T}$ being the time-ordering operator. Thus, $\hat{U}_T$ defines an effective static system $\hat{H}_\text{eff}=iT^{-1}\ln\hat{U}_T$ whose energy spectrum matches with the quasienergy spectrum of our periodic system. Then one can use the well-developed tool of topological phases in static systems to study periodic systems via $\hat{H}_\text{eff}$. Applying the Floquet theorem on our Hamiltonian, we have
\begin{equation} \label{FD}
\mathcal{H}_{\textrm{eff}}(\textbf{k})=iT^{-1}\,\textrm{ln}\,[e^{-i\mathcal{H}_{2}(\textbf{k})T_{2}}e^{-i\mathcal{H}_{1}(\textbf{k})T_{1}}],
\end{equation}
where $\mathcal{H}_{j}(\textbf{k})$ is the Hamiltonian with $m$ replaced by $m_j$.

Different from the static case, the topological phases of a periodic system can be carried by a quasienergy gap not only at zero but also at $\pi/T$, which makes the topological description in static systems insufficient. We can establish a complete topological description of our periodic system from $\mathcal{H}_\text{eff}({\bf k})$. The first-order topology is described by the winding number, which requires the chiral symmetries. However, $\mathcal{H}_{\textrm{eff}}(\textbf{k})$ does not inherit the chiral symmetry of the static system due to $[\mathcal{H}_{1}({\bf k}), \mathcal{H}_{2}({\bf k})]\neq 0$. We make two unitary transformations $G_{l}(\textbf{k})=e^{i(-1)^{l}\mathcal{H}_{l}(\textbf{k})T_{l}/2}$ ($l=1,2$), which do not change the quasienergy spectrum, to recover the chiral symmetry and obtain $\mathcal{\widetilde{H}}_{\textrm{eff},l}(\textbf{k}) = iT^{-1}\ln[{G_{l}(\textbf{k}){U}_{T}(\textbf{k})G_{l}^{\dag}(\textbf{k})}]$ \cite{PhysRevB.90.125143,PhysRevB.103.L041115}. Then two winding numbers $W_l$ defined in $\mathcal{\widetilde{H}}_{\textrm{eff},l}(\textbf{k})$ relate to the first-order topologies of $\mathcal{H}_{\textrm{eff}}(\textbf{k})$ at the quasienergies $\alpha/T$, with $\alpha=0$ or $\pi$, as
\begin{equation}
W_{\alpha/T}=(W_{1}+e^{i\alpha}W_{2})/2.
\end{equation} The number of $\alpha/T$-mode drumhead surface states is equal to $2|W_{\alpha/T}|$. Being the same as the static case, the second-order topology is captured by the quadrupole moment.

The quasienergy spectrum of $\mathcal{H}_{\textrm{eff}}({\bf k})$ in Fig. \ref{fig:3}(a) reveals that the second-order corner states are present at the quasienergy zero, which is witnessed by the nontrivial $P$ in Fig. \ref{fig:3}(b). We find that large parts of the quasienergy gaps at both of zero and $\pi/T$ are closed. The winding numbers $W_{\alpha/T}$ in Figs. \ref{fig:3}(c) and \ref{fig:3}(d) reveal that the regimes with closed quasienergy gaps exhibit more colorful first-order topological phases than the static case in Fig. \ref{fig:2}(b). In particular, the large-winding-number phases with $W_{\pi/T}=\pm2$ and thus the enhanced numbers of the drumhead surface states, which are absent in the static case, are present. Combined with the second-order topological phases in Fig. \ref{fig:3}(b), this result indicates that a hybrid-order nodal-line semimetal, with coexisting second-order nodal lines in the zero mode and first-order ones in the $\pi/T$ mode, are created by periodic driving. Figures \ref{fig:3}(e) and \ref{fig:3}(f) show the distribution of the zero- and $\pi/T$-mode nodal lines in the Brillouin zone. We see that the nodal-line structures are dramatically changed by periodic driving compared with the static case in Fig. \ref{fig:2}(c). First, the separated nodal loops in the $k_x=0$ plane of Fig. \ref{fig:2}(c) are merged. Second, several nodal chains, where the nodal loops in the two planes of $k_x=0$ and $\pi$ are connected by nodal lines, are formed in the zero mode, and a crossing ring nodal net, where the three nodal loops in the planes of $k_x=0$ and $k_y=0$ are linked together, are formed in the $\pi/T$ mode. We thus realize an exotic hybrid-order nodal-line semimetal, with coexisting second-order nodal chains and first-order crossing ring nodal net. Such rich nodal-line structures are difficult to realize in static systems. The hybrid-order nodal-line topological semimetal is confirmed by the probability distributions of the zero- and $\pi/T$-mode states. The second-order hinge Fermi arcs and the first-order drumhead surface states separated by the nodal lines in Fig. \ref{fig:3}(g) verify the second-order nature of the zero-mode nodal lines. The purely drumhead surface states in Fig. \ref{fig:3}(h) confirm the first-order nature of the $\pi/T$-mode nodal lines. Therefore, our result reveals that periodic driving supplies an efficient tool to adjust the nodal-line structure and the topological phases of the SONLS.

\begin{figure}[tbp]
\centering
\includegraphics[width=\columnwidth]{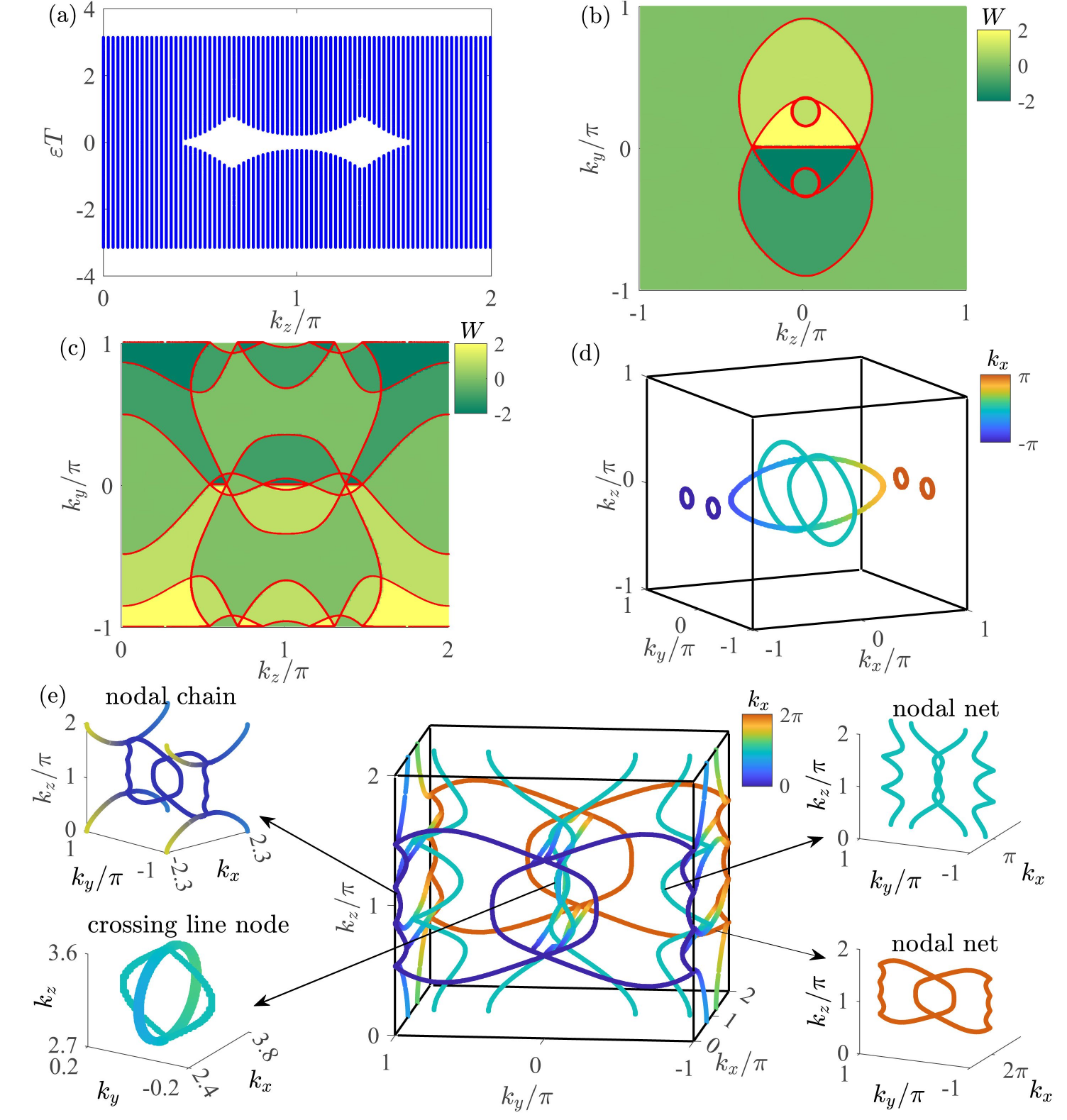}
\caption{ (a) Quasinergy spectrum and winding numbers (b) $W_0$ and (c) $W_{\pi/T}$ of the periodically driven system. Nodal-line distribution of the (d) zero-mode and (e) $\pi/T$-mode in the Brillouin zone. We use $\gamma=0.5f$, $a=0.6f$, $\lambda=0.3f$, $m_{1}=0$, $m_{2}=1.1$, $T_{1}=2.8f^{-1}$, and $T_{2}=1.2f^{-1}$.}\label{fig:4}
\end{figure}

Furthermore, periodic driving can also realize the conversion of different orders and structures of nodal lines, both of which are hard to realize in static systems. The quasienergy spectrum in Fig. \ref{fig:4}(a) shows that the zero-mode corner states disappear and the $\pi/T$-mode gap is persistently closed. Widely changing from $-2$ to $2$, $W_{\alpha/T}$ in Figs. \ref{fig:4}(b) and \ref{fig:4}(c) reveals rich first-order topological phases in the gap closing regimes both for the zero and $\pi/T$ modes. Thus the numbers of drumhead surface states are enhanced. It is interesting to see that fruitful nodal-line structures are created. The zero mode shows a crossing ring nodal net, [see Fig. \ref{fig:4}(d)]. The $\pi/T$ mode shows a nodal chain, a crossing line node \cite{PhysRevB.95.245208}, and two nodal nets [see Fig. \ref{fig:4}(e)]. Both of the zero- and $\pi/T$-mode states are first-order drumhead surface states. The results verify that the SONLS in Fig. \ref{fig:2} is converted into a first-order nodal-line semimetal with rich topological phases and nodal-line structures both at the zero and $\pi/T$ modes by periodic driving.

\section{Discussion and conclusion}
The step like driving protocol is considered merely for the convenience of numerical calculation. Our scheme is generalizable to other driving forms, which can be discretized into a finite number of step functions. A second-order topological semimetal has been predicted in Cd$_3$As$_2$, KMgBi, and PtO$_2$ \cite{PhysRevLett.124.156601,RN5}, and realized in acoustic metamaterials \cite{RN4,Wei2021}. Floquet engineering has exhibited its ability in creating exotic phases in electronic material \cite{Mahmood2016,McIver2020}, ultracold atom \cite{Wintersperger2020}, superconductor qubit \cite{Roushan2017}, and photonic \cite{RN6,Mukherjee2017,Maczewsky2017,PhysRevLett.122.173901} systems. Different topological semimetals have been realized in photonic \cite{Chen2016,Yang2017,Yan2018,doi:10.1126/science.aaa9273}, ultracold-atom \cite{Song2019,doi:10.1126/science.abc0105}, and circuit-QED \cite{Mei_2016} systems. This progress indicates that our prediction is realizable in state-of-the-art quantum-simulation platforms.

In summary, we have proposed a scheme to generate SONLS in a $\mathcal{PT}$-symmetry broken system and explored its diverse variations in both nodal-line structures and topological orders by Floquet engineering. An exotic hybrid-order nodal-line semimetal and abundant nodal-line structures including nodal chains, crossing ring nodal nets, crossing line nodes, and nodal nets are created by periodic driving. This result enriches the classification of topological semimetals and provides a convenient way to reduce the practical difficulties in adjusting the nodal-line structures, Fermi arcs, and drumhead surface states in static systems. Such an on-demand controllability to the topological semimetal is helpful in designing novel topological devices by simultaneously utilizing the advantages of two topological orders and various nodal-line structures of the semimetal.
\section*{Acknowledgments}
The work is supported by the National Natural Science Foundation of China (Grants No. 12275109, No. 11875150, and No. 12247101) and the Supercomputing Center of Lanzhou University.

\bibliography{references}

\end{document}